\documentclass[journal]{IEEEtran}
\usepackage{fontenc}
\usepackage{multirow}
\usepackage{cellspace}
\usepackage{caption}
\usepackage{stfloats}
\usepackage{makecell}
\usepackage[colorlinks,
linkcolor=black,
anchorcolor=black,
citecolor=black]{hyperref}
\usepackage{amsfonts}
\usepackage{mathrsfs}
\usepackage{amsfonts}
\usepackage{ifpdf}
\usepackage[justification=centering]{caption}
\usepackage{cite}
\ifCLASSINFOpdf
\usepackage[pdftex]{graphicx}
\else \fi
\usepackage[cmex10]{amsmath}
\usepackage{amsmath,amsfonts}
\usepackage{array}
\usepackage{mdwmath}
\usepackage{mdwtab}
\usepackage{eqparbox}
\usepackage[tight,footnotesize]{subfigure}
\usepackage{algorithm}
\usepackage{algorithmicx}
\usepackage{algpseudocode}
\usepackage{amsmath}
\usepackage{epsfig}
\usepackage{ae}
\usepackage{amssymb}
\usepackage{times}  
\usepackage{booktabs}
\usepackage{color}
\usepackage{bm}

\usepackage{epstopdf}
\usepackage{stfloats}
\usepackage{subfigure}
\usepackage{float}
\usepackage{url}
\usepackage{threeparttable}

\floatname{algorithm}{Algorithm}

\begin{document}
	\title{Decomposition Model Assisted Energy-Saving Design in Radio Access Network}
	
	\author{Xiaoxue~Zhao,
		Yijun~Yu,
		Yexing~Li,
		Dong~Li,
		Yao~Wang,~\IEEEmembership{Graduate Student Member,~IEEE}
		and~Chungang~Yang,~\IEEEmembership{Senior Member,~IEEE}
		\vspace{-1.5em}
		\thanks{This work was supported by a cooperative project from Huawei Technologies Company Ltd. (TC20221114023).}
		\thanks{Xiaoxue Zhao, Yao Wang, and Chungang Yang are with the State Key Laboratory of Integrated Services Networks, Xidian University, Xi'an, 710071, China (email: zhaoxiaoxue1123@163.com; chgyang2010@163.com; yaow0518@gmail.com).}
		\thanks{Yijun Yu is with the Research Department of Autonomous Driving Network, Huawei Technologies Company Ltd., Dongguan, 523808, China (e-mail: yuyijun@huawei.com).}
		\thanks{Yexing Li, and Dong Li are with the Research Department of Autonomous Driving Network, Huawei Technologies Company Ltd., Shanghai, 201206, China (e-mail: liyexing@huawei.com; layton.lidong@huawei.com).}
	}\maketitle

\begin{abstract}
	The continuous emergence of novel services and massive connections involve huge energy consumption towards ultra-dense radio access networks. Moreover, there exist much more number of controllable parameters that can be adjusted to reduce the energy consumption from a network-wide perspective. However, a network-level energy-saving intent usually contains multiple network objectives and constraints. Therefore, it is critical to decompose a network-level energy-saving intent into multiple levels of configurated operations from a top-down refinement perspective. In this work, we utilize a softgoal interdependency graph decomposition model to assist energy-saving scheme design. Meanwhile, we propose an energy-saving approach based on deep Q-network, which achieve a better trade-off among the energy consumption, the throughput, and the first packet delay. In addition, we illustrate how the decomposition model can assist in making energy-saving decisions. Evaluation results demonstrate the performance gain of the proposed scheme in accelerating the model training process. 
	
\end{abstract}

\begin{IEEEkeywords} Radio access networks, energy-saving intent, decomposition model, deep Q-network, softgoal interdependency graph
\end{IEEEkeywords}

\section{Introduction}
\IEEEPARstart{T}{owards} 5th generation (5G) networks, there are various types of communication services and massive terminal connections lead to a significant increase in the energy consumption of base stations (BSs) [1]. The 6th generation (6G) networks, On the one hand, face more stringent service requirements including higher user experience rates and very low delay. On the other hand, the deployment scale of BSs will be several times larger than that of 5G to achieve coverage comparable to that of 5G [2]. This means that 6G networks will face huger energy consumption demands [3]. Therefore, to cope with the challenge of high energy consumption and promote the green and low-carbon 6G development, reducing the energy consumption of BSs has become urgent [4].

However, energy-saving design is a complex problem because there are multiple coupled network objectives with huge number of adjustable parameters from a network-wide perspective. Based on our previous project, on the one hand, we found that the energy-saving intent is coupled with multiple network objectives such as throughput and delay. Reducing energy consumption may result in a decrease in throughput along with an increase in delay. On the other hand, there are numerous parameters that can be adjusted in BSs to reduce energy consumption, such as transmit power and antenna downtilt angle. In addition, the increase in the number of network elements makes the problem of dimensional explosion for energy-saving operations. Therefore, it is a challenge to effectively balance the network objectives to ensure the continuous realization and optimization of energy-saving intent.

Recent works have explored the issue of energy-saving optimization in radio access networks domain and have also achieved some results. The authors in [5] transformed the  energy-saving problem into a multi-parameter multi-objective optimization problem and constructed a multi-objective solution model based on genetic algorithms. Similarly, the authors in  [12]-[14] also transformed the energy-saving problem into a multi-objective optimization problem to solve. User equipments (UEs) continuously gather a substantial data via radio measurements [6], providing rich application scenarios for reinforcement learning [7]. The works [15]-[19] used reinforcement learning for energy-saving optimization decisions to explore the trade-off between network objectives among energy consumption. Additionally, the authors in [8] and [9] introduced artificial intelligence techniques to predict and optimize the management of energy consumption. However, few of these works have consider the potential conflicts between energy-saving intent and other network objectives from a network-level perspective, ignoring the need to construct decomposition model. Meanwhile, there is a lack of solutions that explore the interaction between the decomposition model and the RL model from the decomposition model assisted energy-saving scheme.

In this paper, we present a decomposition model to assist energy-saving design, and our framework is divided into the design time and the run time. During the design time, we establish a decomposition model based on softgoal interdependency graph, and fully consider the conflicts between energy-saving intent and network objectives in the decomposition process. Ultimately, we decompose the energy-saving intent into specific and executable energy-saving operations from top to down [10]. Combining the concepts of convolutional neural networks with Q-learning, the deep Q-network (DQN) is able to effectively deal with the multidimensional discrete operation space and learn effective energy-saving operations from the complex environment [11]. To improve the accuracy of the DQN in making decisions, during the run time, we design the scheme for the decomposition model to assist the DQN in making energy-saving decisions. The scheme also includes the methods for updating weights in the decomposition model and identifying conflicts in the decomposition process. Finally, we conduct evaluation to validate the scheme, and the results demonstrate that the scheme can effectively trade-off among the energy consumption, the throughput, and the first packet delay. 

The goal of this paper is to explore efficient energy-saving operations to achieve the network-level energy-saving intent. This paper investigates the problem of decomposition model assisted energy-saving design in radio access networks scenario. The novel contributions of this paper are summarized as follows:

\begin{itemize}
	\item[$\bullet$]  We present a decomposition model assisted energy-saving design scheme with the capacities to update the weights and identify the conflicts.
	
	\item[$\bullet$] We propose a DQN-based method for choosing energy-saving operations assisted by the decomposition model, efficiently trade-off among the energy consumption, the throughput and the first packet delay.
	
	\item[$\bullet$] We implement a proof of concept evaluation to show the effectiveness of our proposed decomposition model assisted energy-saving scheme.

\end{itemize}

The rest of this paper is organized as follows. In Section II, we provide a review of related works. Section III describes the problem formulation. In Section IV, we introduce the specific flow of the decomposition model assisted energy-saving scheme, and introduce methods for updating weights and identifying conflicts. Section V describes the realization of the scheme. Finally, we provide the simulation results in Section VI, and conclude our work in Section VII.

\section{Related Works}
\subsection{Multi-Objective Optimization to Solve Energy-Saving Problem}\label{A}
The issue of energy-saving in RAN scenario is a significant challenge. The current methods consider the energy-saving intent and network objectives correlation [12]-[14]. The energy-saving problem is presented as a multi-objective optimization problem and solved using an optimization algorithm. However, most of these methods do not further explore the complex coupling relationship between the energy-saving intent and other network objectives. 

The authors in [12] investigated the issue of user association to optimize energy efficiency while adhering to a specific spectral efficiency objective. In order to address the challenge of associating users with energy and spectrum efficiency, they established an optimization model that they then turned into a quadratic assignment problem. They solved the problem via a heuristic algorithm. In [13], the authors established the optimization objective to satisfy the throughput demand while selecting the appropriate number of service cells such that the total energy consumption is minimized. Then, the authors proposed a centralised user association strategy and a switching algorithm based on neighbouring cell loads to simplify the solution. In [14], the authors considered heterogeneous scenarios, using long and short-memory to forecast the services of Micro BSs. They implemented a strategy to indentify and deactive low-load Micro BSs. Additionally, a genetic algorithm is used to solve the well established multi-objective optimization model of BSs.

The above works only establish a multi-objective optimization problem and simplify the problem to solve it, which is not sufficient to cope with network-level energy-saving intent. This is due to that the network-level energy-saving intent usually contains multiple network objectives or performance constraints. Therefore, it is necessary to establish a decomposition model. In this decomposition process, the constraints between energy-saving intent and network objectives are considered to ensure that energy-saving is pursued without compromising other network objectives.

\subsection{Reinforcement Learning to Solve Energy-Saving Problem}\label{B}

The current solutions are mainly based on reinforcement learning for multi-level sleep models management to optimize energy-saving intent [15]-[19]. There are few solutions that take the perspective of a decomposition model assisted energy-saving scheme, and the interaction between a decomposition model and a reinforcement learning model has rarely been investigated. As a result, the improvement in energy-saving optimization that can be achieved by combining the two needs to be further explored.

The authors in [15] considered multi-level sleep models and introduced a reinforcement learning algorithm specifically designed for small cells. This system adjusted the activities of the small cells to meet a delay limit. The algorithm utilized cochannel interference, the buffer capacity of the BSs, and the anticipated throughput to intelligently determine the optimal policy for the sleep models. A management method using Q-learning was proposed in [16] to optimize the trade-off between energy consumption and delay. It also determined the optimal length of different sleep models for the BSs. The authors in [17] found the trade-off between energy consumption and quality of service limitations by using state action reward state action (SARSA) combined with multi-level management of sleep models. In [18], the authors considered four different sleep models, each characterized by a unique activation time and energy consumption. Then an adaptive algorithm based on SARSA was proposed to realize the optimal trade-off between delay and energy consumption, which could accommodate the selection of the sleep models for the BSs. To describe various optimization techniques in a decentralized way, the authors in [19] presented a network energy management framework based on multi-policy multi-objective hierarchical reinforcement learning. The framework employed distributed double deep Q-network (DDQN) to optimize energy consumption and throughput.

In wireless environments, energy-saving of BSs is intricate due to its dynamic characteristics, numerous performance metrics and complex interdependence. Therefore, it is essential to establish a decomposition model for assisting energy-saving design, which decomposing the network-level energy-saving intent and supporting the energy-saving decision-making process. In this study, we present a decomposition model assisted energy-saving design scheme. In particular, the effectiveness of the DQN assisted by the decomposition model in the decision-making process is explored.

\begin{figure}[!t]
	\centering
	\includegraphics[width=2.85in]{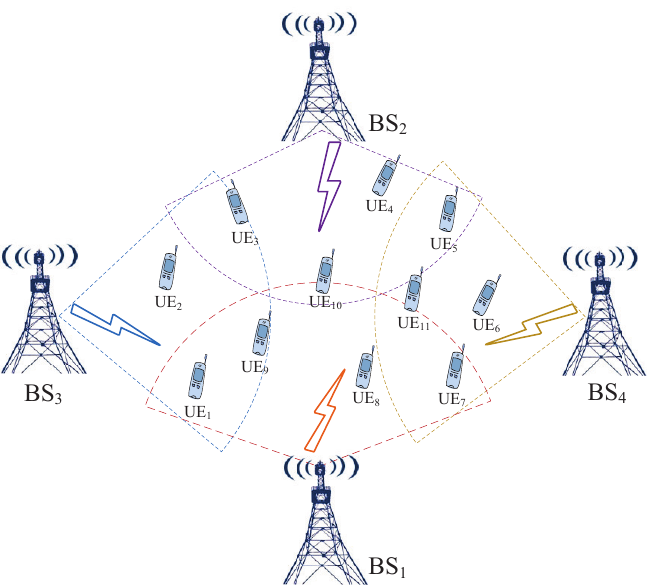}
	\caption{5G radio access networks scenario.}
	\label{Fig1}
	\vspace{-0.45cm}
\end{figure}

\section{Problem Formulation}
\subsection{System Overview}\label{A}
We consider a RAN scenario network with the network size of $N$ BSs and $M$ UEs, as illustrated in Fig. \ref{Fig1}. The set of BSs as ${\cal N} \buildrel \Delta \over = \{ 1,2,...,N\},\forall n \in {\cal N}$, $n$ denotes BS $n$. The UEs are randomly distributed around the BSs, and the set of UEs is defined as ${\cal M} \buildrel \Delta \over = \{ 1,2,...,M\},\forall m \in {\cal M}$, $m$ denotes UE $m$. To simplify, the location of each entity is depicted using a 3D Cartesian coordinate system. The location of  BS $n$ and UE $m$ are denoted as $(x_n^{},y_n^{})$ and  $(x_m^{},y_m^{})$, respectively. Furthermore, each BS is divided into ${N_s}$ sectors, and the altitude is $H$. The system bandwidth of each BS is $B$. The maximum number of physical resource blocks (RBs) can be allocated by BS $n$, denoted as $r_n^{\max } \approx {B \mathord{\left/{\vphantom {B{{B_{prb}}}}}\right.\kern-\nulldelimiterspace} {{B_{prb}}}}$, where ${B_{prb}}$ is the bandwidth of each RB.

Initially, the UEs select and register in the BSs, and discretely generate and add downlink communication services among UEs to the buffer of the BSs. We assume that the service flow of UE $m$ obeys a poisson distribution characterized by an arrival rate of $\lambda$, expressed as $P(X = {\chi _m}) = {{{e^{ - \lambda }}{\lambda ^{{\chi _m}}}} \mathord{\left/
		{\vphantom {{{e^{ - \lambda }}{\lambda ^{{\chi _m}}}} {{\chi _m}!}}} \right.\kern-\nulldelimiterspace} {{\chi _m}!}}$, where ${\chi _m}$ is the number of arriving packets for UE $m$. Different service arrival rates can intuitively reflect the busyness of the services. We assume that the packets size of UE $m$ is ${M_m}$ bits, and the time interval among the arrival of neighboring packets ${t_m}$ obeys the exponential distribution with expectation value $1/\lambda $, which can be expressed as $f(t_m^{}) = \lambda  * {e^{ - \lambda t_m^{}}}$.

When the service packets of the UEs arrive at the BSs, the BSs start to perform resource scheduling, i.e., assigning the RBs to the UEs. Specifically, in each transmission time interval (TTI), the UEs maps the channel quality indicator (CQI) of the downlink based on signal to interference plus noise ratio (SINR), and reports the CQI to the BSs. The set of CQI numbers is represented as ${\cal K} \buildrel \Delta \over = \{ 1,2,...,K\},\forall k \in {\cal K}$, $k$ denotes CQI $k$. BS $n$ determines the modulation and coding scheme (MCS) after receiving CQI $k$ reported by UE $m$. The maximum code rate of UE $m$ at CQI $k$ is specified by the protocol, and denoted as $Cr_{m,k}^{\max }$. The number of the RBs to be assigned to UE $m$ is determined based on the packets size ${M_m}$, and the code rate $Cr_{m,k}^{}$ is further calculated. Therein, $Cr_{m,k}^{}$ is not exceeded $Cr_{m,k}^{\max }$, the resource scheduling is performed. Otherwise, the MCS indicator is lowered by one order until ${Cr_{m,k}^{} \le Cr_{m,k}^{\max }}$, and then the resource scheduling is performed. It can be expressed by the following equation

\vspace{-0.2cm}
\begin{equation}
	\label{deqn_ex1}
	\left\{ {\begin{array}{*{20}{c}}
			{Cr_{m,k}^{} \le Cr_{m,k}^{\max },}&{{\rm{Perform\; scheduling}}},\\
			{Cr_{m,k}^{} > Cr_{m,k}^{\max },}&{{\rm{Decrease\; MCS\; indicator }}}.
	\end{array}} \right.
\end{equation}
\vspace{-0.5cm}

\subsection{Energy Consumption Model}\label{B}
The energy consumption of each BS is not only strongly correlated with the transmit power, but also more sensitive to the load [20]. We assume that the power consumption of functional blocks (e.g., cooling) common to all sectors of the BSs is proportional to the size of ${N_s}$. Therefore, the energy consumption of BS $n$ can be expressed as 
\vspace{0.0cm}
\begin{equation}
	\label{deqn_ex1}
	{P_n^{} = (1 - {\zeta _n}){\eta _n}P_{om}^n + {\zeta _n}P_{om}^n},
\end{equation}
\vspace{-0.4cm}

\noindent
where ${\zeta _n}$ is a constant, $0 \le {\zeta _n} \le 1$. ${\zeta _n}=1$ is the constant power consumption model, ${\zeta _n} = 0$ is the full power consumption proportionality model, and $0 < {\zeta _n} < 1$ is the non-power consumption proportionality model. $P_{om}^n$ is the maximum operating power fully utilized by BS $n$, denoted as $P_{om}^n = {\xi _n}P_{Tx}^n + {\psi _n}$, where $P_{Tx}^n$ is the transmit power of BS $n$, ${\xi _n}$ and ${\psi _n}$ are constants. Without lose the generality, we assume that $\forall n \in {\cal N}$, there are ${\zeta _n} = \zeta$, ${\xi _n} = \xi$, and ${\psi _n} = \psi$. Thus, the load of BS $n$ measured by the RBs utilization can be expressed as

\vspace{-0.25cm}
\begin{equation}
	\label{deqn_ex1}
	\setlength\abovedisplayskip{6pt}
	\setlength\belowdisplayskip{6pt}	
	{\eta _n} = \frac{{\sum\nolimits_{m \in {\cal M}}^{} {{\delta _{n,m}}{r_{n,m}}} }}{{{r_n}}},
\end{equation}
\vspace{-0.3cm}

\noindent
where $0 \le {\eta _n} \le 1$, and ${r_{n,m}}$ is the number of the RBs allocated by BS $n$ to UE $m$. ${\delta _{n,m}}$ is the connection between UE $m$ and BS $n$. If ${\delta _{n,m}} = 1$, UE $m$ is connected to BS $n$. If ${\delta _{n,m}} = 0$, UE $m$ is not connected to BS $n$.
\vspace{-0.3cm}

\subsection{Downlink Communication Model}\label{C}
For measuring the throughput when the BSs establish a downlink with the UEs in the cell, we assume that the access mode between the BSs and the UEs is orthogonal frequency-division multiplexing (OFDM) mode, and the UEs can obtain the accurate channel state information (CSI) [21]. The distance between BS $n$ and UE $m$ can be calculated by $d_{n,m}^{} = \sqrt {{{\left( {x_n^{} - x_m^{}} \right)}^2} + {{\left( {y_n^{} - y_m^{}} \right)}^2} + {H^2}}$. We select a 3DUmaLOS line-of-sight transmission model [22], and the path loss between UE $m$ and BS $n$ can be given as

\vspace{-0.3cm}
\begin{equation}
	\label{deqn_ex1}
	PL_{n,m}^{} = 28.0 + 22{\log _{10}}\left( {{d_{n,m}}} \right) + 20{\log _{10}}\left( {{f_c}} \right),
\end{equation}
\vspace{-0.4cm}

\noindent
where $f_c^{}$ is the system frequency. The channel gain between BS $n$ and UE $m$ can be expressed as $g_{n,m}^{} = g_a^n - PL_{n,m}^{} - \alpha$, where $g_a^n = 10 - 20\lg (\cos \frac{{{\theta _n}\pi }}{{180}})$ is the antenna gain of BS $n$, ${\theta _n}$ is the angle adjustment of the antenna downtilt for BS $n$, and $\alpha$ is the shadow attenuation. Since the signal received by the UEs from the serving BSs is regarded as the target signal, while the signals from other BSs are considered as the interference signals. Thus, the SINR between BS $n$ and UE $m$ is denoted as $SINR_{n,m}^{} = \sum\nolimits_{i \in {\cal N},i \ne n} {P_{Tx}^ng_{i,m}^{}}$. Therefore, the throughput between BS $n$ and UE $m$ is denoted as

\vspace{-0.2cm}
\begin{equation}
	\label{deqn_ex1}
	C_{n,m}^{} = r_{n,m}^{}{B_{prb}}\log _2^{}(1 + \frac{{P_{Tx}^ng_{n,m}^{}}}{{\sum\limits_{i = 1,i \ne n}^N {P_{Tx}^ng_{i,m}^{}}  + \sigma _{}^2}}),
\end{equation}
\vspace{-0.3cm}

\noindent
where $\sigma _{}^2$ is the noise power.
\vspace{-0.2cm}

\subsection{First Packet Delay Model}\label{D}
We calculate the first packet delay from the transmission delay and the queuing delay statistics. The service packets of the UEs arrive at the buffer of the BSs, and the BSs adhere to the principle of \textit{first arriving first serving} to process the packets. The service packets arriving later need to queue up in the buffer to wait for the processing. Thus, the queuing delay $T_{queue}^m$ refers to the packets from the time when it enters the buffer of the BSs to the difference between the time when it leaves the buffer. The transmission delay $T_{trans}^m$ is related to the packets size $M_m$ and the link bandwidth ${{r_{n,m}}R_{prb}^{m,k}Cr_{m,k}^{}}$, where $R_{prb}^{m,k}$ is the rate of each RB. Therefore, the first packet delay between BS $n$ and UE $m$ can be determined as

\vspace{0.0cm}
\begin{equation}
	\renewcommand{\arraystretch}{1} 
	\label{deqn_ex1}
	\begin{array}{l}
		T_{ave}^m = T_{queue}^m + T_{trans}^m\\[8pt]
		\;\;\;\;\;\;\;{\rm{    }} = (T_{out}^{n,m} - T_{in}^{n,m}) + \frac{{{M_m}}}{{{r_{n,m}}R_{prb}^{m,k}Cr_{m,k}^{}}},
	\end{array}
\end{equation}
\vspace{-0.1cm}

\noindent
where $T_{in}^{n,m}$ is the time when the service packets of UE $m$ is added to the buffer of BS $n$, and $T_{out}^{n,m}$ is the time when the service packets of UE $m$ leaves the buffer of BS $n$.

\subsection{Optimization Problem}\label{D} 
We aim to improve the throughput and reduce the first packet delay while reducing the energy consumption by adopting energy-saving operations. Therefore, the optimization problem is mathematically expressed as
\vspace{0.1cm}
\begin{equation}
	\label{deqn_ex1}
	\begin{array}{l}
		\!\!\!\!\!\!\!\!\!\!\!\!\!\!\!\!\!\!\!\!\!\!\!\!\!\!\!\!\!\!\!\!\!\!\!\!\!\!\!\!\!\!\!\!\!\!\!\!\min \;\;\;\;\;\sum\limits_n {(1 - {\zeta _n}){\eta _n}P_{om}^n + {\zeta _n}P_{om}^n} 
	\end{array}
\end{equation} 
\vspace{-0.1cm}
\begin{equation}
	\label{deqn_ex1}
	\begin{array}{l}
		\begin{array}{*{20}{c}}
			{{\rm{s}}{\rm{.t}}{\rm{.\;\; }}{C_1}{\rm{:}}}&{P_{Tx}^n \le P_{Tx,\max }^n},\forall n \in {\cal N},
		\end{array}\\[8pt]
		\begin{array}{*{20}{c}}
			{\;\;\;}&{{C_2}:}&{{r_{n,m}} \le {r_n},\forall m \in {\cal M}},
		\end{array}\\[8pt]
		\begin{array}{*{20}{c}}
			{\;\;\;}&{{C_3}:}&{P_{Tx}^ng_{n,m}^{} > {P_0},\forall m \in {\cal M},\forall n \in {\cal N},}
		\end{array}\\[8pt]
		\begin{array}{*{20}{c}}
			{\;\;\;}&{{C_4}:}&{\sum\limits_{n = 1}^N {P_n^{} \le P_{\max }^{}} },\forall n \in {\cal N},
		\end{array}\\[10pt]
		\begin{array}{*{20}{c}}
			{\;\;\;}&{{C_5}:}&{\sum\limits_{m = 1}^M {{\delta _{n,m}}C_{n,m}^{}} }\ge {C_{\min }},\forall m \in {\cal M},\forall n \in {\cal N},
		\end{array}\\[11pt]
		\begin{array}{*{20}{c}}
			{\;\;\;}&{{C_6}:}&{\sum\limits_{m = 1}^M {{\delta _{n,m}}T_{ave}^m}  \le {T_{\min }}},
		\end{array}
	\end{array}\nonumber
\end{equation} 
\vspace{0.0cm}

\noindent
where $P_{Tx,\max }^n$, ${P_{\max }^{}}$, and ${T_{\max }^{}}$ are the maximum of the transmit power, the energy consumption, and the first packet delay, respectively. ${C_{\min }^{}}$ is the minimum of the throughput. $C_1$ is to limit the BS transmit power not to exceed the maximum transmit power $P_{Tx,\max }^n$. $C_2$ indicates that the number of RBs assigned by the BS to each UE cannot exceed the number of RBs that can be assigned. $C_3$ shows that the signal power received by each UE cannot be lower than ${P_0}$, otherwise it cannot be guaranteed that the UE can receive the signal correctly. $C_4$, $C_5$ and $C_6$ are used to ensure the performance requirements for the energy consumption, the throughput, and the first packet delay are achieved, respectively.

\section{Decomposition Model Assisted Energy-Saving Scheme}
To solve the formulated optimization problem, we introduce the implementation flow of the decomposition model assisted energy-saving scheme, and explain how the design time and the run time can realize effective association and collaboration. Then we introduce the methods of updating weights and identifying conflicts, respectively.

\subsection{Basic Idea}\label{A}

\begin{figure}[!t]
	\centering
	\includegraphics[width=2.5in]{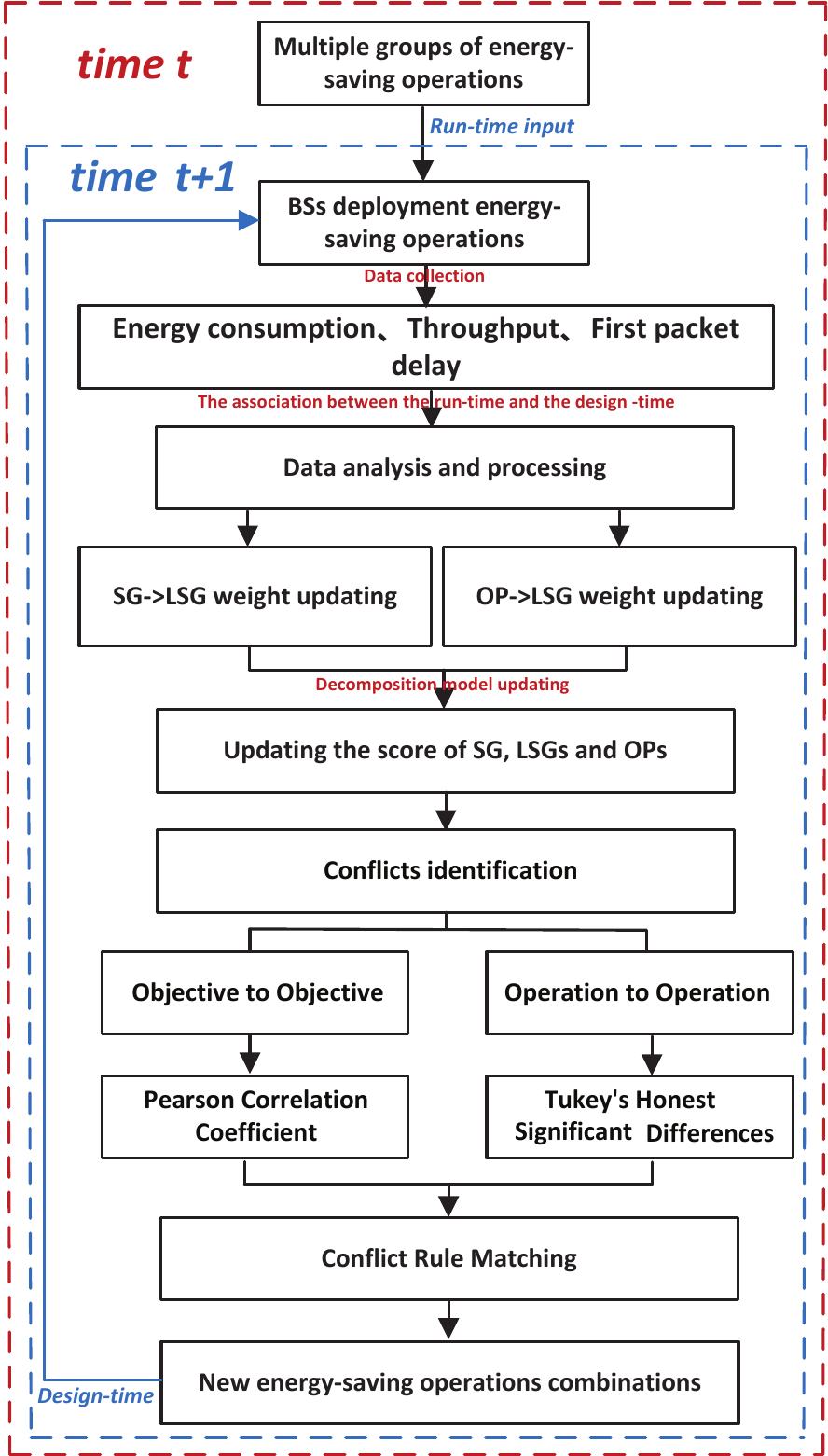}
	\caption{The specific implementation flow of the decomposition model assisted energy-saving scheme.}
	\label{Fig2}
	\vspace{-0.45cm}
\end{figure}

As illustrated in Fig. 2, we describe the specific implementation flow of the decomposition model assisted energy-saving scheme. We consider three energy-saving operations, the adjustment of the transmit power, the adjustment of the antenna downtilt angle, and the adjustment of the BSs sleep. At the moment $t$, the decomposition model outputs multiple combinations of energy-saving operations at different granularities, which are stored as JSON format, as shown in Fig. 3. This JSON file is passed to the run time as the input and constitutes the action space of the DQN. Then, the BSs agent interacts with the simulation environment, selects a combination of energy-saving operations from the action space and executes them. This results in changes among the energy consumption, the throughput, and the first packet delay. We collect data on these network objectives, analyze and process the data during the design time. The decomposition model contains three components, softgoal (SG) and leaf-softgoals (LSGs), and operations (OPs), where SG represents energy-saving intent, LSGs represent network objectives and OPs represent energy-saving operations [10]. Firstly, we utilize methods to update the weights between SG and LSGs, and the weights between LSGs and OPs. Secondly, we recalculate the scores of the energy-saving intent, all LSGs and all OPs using the formulas in [23]. Then, we identify the conflicts among network objectives, and the conflicts among energy-saving operations, respectively. The process of conflict identification is matched with predefined conflict rules. The matching output will be used as a filtering condition to exclude the unqualified energy-saving operations from all the energy-saving operations combinations, and finally output the new energy-saving operations combinations. At the moment $t+1$, the new combinations will be passed to the DQN again.

\begin{figure}[!t]
	\centering
	\includegraphics[width=2.4in]{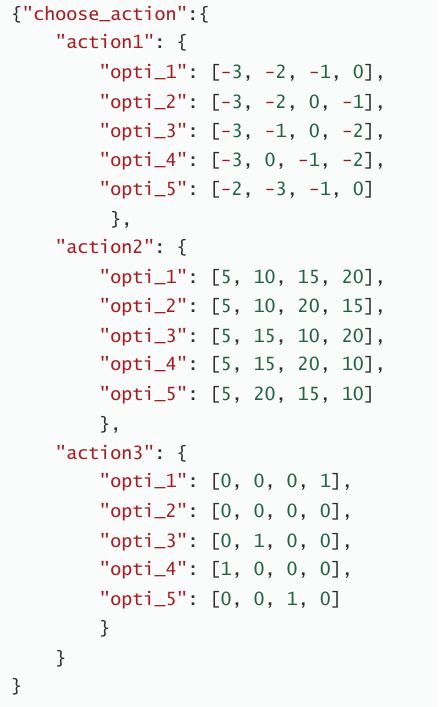}
	\caption{The combinations of the three energy-saving operations at different granularities are stored as JSON format.}
	\label{Fig3}
	\vspace{-0.45cm}
\end{figure}

In summary, the decomposition model assisted energy-saving scheme relies on the association coupling between the design time and the run time. The design time outputs the energy-saving operations combinations, and the BSs agent of the run time verifies and selects the optimal energy-saving operations for each BS. The network objectives data of the run time is transmitted to the design time in real time, which assists the design time to change the output energy-saving operations combinations in real time. The decomposition model can effectively improve the action space of the DQN by identifying conflicts between energy-saving operations and formulating conflict rules, which enables the DQN to explore the effective energy-saving operation space and improve the accuracy of decision-making.

\subsection{Weights Update}\label{B} 
There are two kinds of weights in the decomposition model:

\begin{itemize}
	\item[$\bullet$] between SG and LSG: this weight reflects the importance of LSG in satisfying SG.
	\item[$\bullet$] between LSG and OP: OP contributes to the realization of LSG, and a single OP can have an impact on one or more LSG, so this weight reflects the extent to which OP contribute to LSG, and furthermore the impact on satisfying SG.
\end{itemize}

Based on the data collected on the energy consumption, the throughput, and the first packet delay in the run time, we design weights update methods for these two weights respectively. The methods are described as follows:

\subsubsection{$SG \to LSGs$ weights update }\label{1)} 
we calculate the average of the network objectives data and use the normalization method to restrict its average to between $[-1, 1]$. Taking energy consumption data as an example, assuming a set of energy consumption data as ${\rm{{\cal Z}}}_{}^{EC} = [z_1^{},z_2^{}, \cdots ,z_L^{}]$, the weights between the SG and the LSG can be calculated by

\vspace{-0.3cm}
\begin{equation}
	\label{deqn_ex1}
	{{\mathop{\rm impact}\nolimits} _{SG \times LSG}} = 2 \times \frac{{\bar Z - min(Z_{}^{EC})}}{{max(Z_{}^{EC}) - min(Z_{}^{EC})}} - 1,
\end{equation}
\vspace{-0.2cm}

\noindent
where ${{\mathop{\rm impact}\nolimits} _{SG \times LSG}}$ denotes the weight between the energy-saving intent and the energy consumption network objective, ${\rm{\cal{\bar Z}}}$ represents the average of this set of energy consumption data, $max(Z_{}^{EC})$ and  $min(Z_{}^{EC})$ represent the maximum and minimum value of this energy consumption data respectively.

\subsubsection{$OPs \to LSGs$ weights update}\label{2)} 
we use the expected and actual values of the energy consumption, the throughput and the first-packet delay as inputs to the feedback control method. Where the expected values represent the desired values of the three network objectives in the energy-saving intent. And the actual values are the values of the data collected in the run time. Then, we calcualte the error between the expected values and the actual values of each network objective. The $sigmoid$ function is used to adjust the error and generate the weight adjustment value. Finally, the weights between the LSG and the OP can be calculated by

\vspace{-0.4cm}
\begin{equation}
	\label{deqn_ex1}
	\begin{array}{l}
		{{\mathop{\rm impact}\nolimits} _{LSG \times OP}} = clip(sigmoid (er),{\alpha _{{{\min }^{}}}},{\alpha _{{{\max }^{}}}})
		
	\end{array}
\end{equation}
\vspace{-0.2cm}
\begin{equation*}
	\label{deqn_ex1}
	\;\;\;\;\;= \left\{ {\begin{array}{*{20}{c}}
			\begin{array}{l}
				\alpha _{\min }^{}\\[6pt]
				sigmoid(er)\\
				\alpha _{\max }^{}
			\end{array}&\begin{array}{l}
				if{\rm{\; }}sigmoid(er) < \alpha _{\min }^{},\\[6pt]
				if{\rm{\; }}\alpha _{\min }^{} \le sigmoid \le \alpha _{\min }^{},\\[6pt]
				if{\rm{ \;}}sigmoid(er) > \alpha _{\max }^{},
			\end{array}
	\end{array}} \right.
\end{equation*}
\vspace{0.0cm}

\noindent
where $er = {\rm{{\cal Z}}}_{desired}^{} - {\rm{{\cal Z}}}_{actual}^{}$, ${\rm{{\cal Z}}}_{desired}^{}$ denotes the expected value of the energy consumption, ${\rm{{\cal Z}}}_{actual}^{}$ denotes the actual value of the energy consumption; $sigmoid(er) = {1 \mathord{\left/
		{\vphantom {1 {(1 + {e^{ - er}})}}} \right.\kern-\nulldelimiterspace} {(1 + {e^{ - er}})}}$ the error value is calculated as a function of  the error value; $\alpha _{\min }^{} =  - 1$, $\alpha _{\max }^{} = 1$.

\subsection{Conflicts Identification}\label{D} 
The network-level energy-saving intent are coupled with network objectives, so conflicts between network objectives need to be considered in the process of decomposition. In addition, conflicts between multiple energy-saving operations are mainly related to mutual interference, inconsistent goals, or different impacts of different operations on performance. Therefore, identifying these two types of conflicts is necessary.

\subsubsection{Objective-to-Objective}\label{1)} 
conflicts among network objectives are a common phenomenon, for example the expectation of reducing energy consumption while increasing throughput, when there is a conflict between these two objectives. Therefore, we use the pearson's correlation coefficient (PPC) to indentify conflicts and evaluate the degree of the conflicts. The PCC is used to quantify the degree of a linear relationship between two variables and is calculated as [25]

\vspace{-0.1cm}
\begin{equation}
	\label{deqn_ex1}
	\rho  = \frac{{\sum\nolimits_{l = 1}^L {(u_l^{} - {\rm{\bar {\cal U}}})(z_l^{} - {\rm{\bar {\cal Z}}})} }}{{\sqrt {\sum\nolimits_{l = 1}^L {(u_l^{} - {\rm{\bar {\cal U}}})_{}^2} } \sqrt {\sum\nolimits_{l = 1}^L {(z_l^{} - {\rm{\bar {\cal Z}}})_{}^2} } }},
\end{equation}
\vspace{-0.2cm}

\noindent
where a set of the throughput data as $\rm{\cal{U}}_{}^{THP} = [u_1^{},u_2^{}, \cdots ,u_L^{}]$, $\cal{\bar U}$ represents the average of this set of throughput data. The conflict level is determined based on the value of $\rho$, which is classified as Low, Medium, and High. $- 1 \le \rho  \le 1$, and the closer the value of $\rho$ is to $-1$, the higher degree of conflict.

\subsubsection{Operation-to-Operation}\label{2)} 
conflicts indentification among multiple energy-saving operations refer to the possible conflicts between different operation granularities under the same energy-saving operation. Therefore, we use tukey's honest significant differences multiple comparison (Tukey's HSD) to identify conflicts [24], and the specific steps are summarized in Algorithm 1. The Student's range statistic $q$ can be determined as

\vspace{0.0cm}
\begin{equation}
	\label{deqn_ex1}
	q = {{{{\bar Q}_1} - {{\bar Q}_2}} \mathord{\left/
			{\vphantom {{{{\bar Q}_1} - {{\bar Q}_2}} {\sqrt {\frac{{OP_{{\mathop{\rm var}} }^{}}}{{O{P_n}}}} }}} \right.
			\kern-\nulldelimiterspace} {\sqrt {\frac{{OP_{{\mathop{\rm var}} }^{}}}{{O{P_n}}}} }},
\end{equation}
\vspace{-0.3cm}

\noindent
where ${\bar Q_1}$ and ${\bar Q_2}$ are the means of the two energy-saving operation combinations data, respectively, $OP_{{\mathop{\rm var}} }^{}$ is the joint variance between the energy-saving operation combinations data, and the number of energy-saving operations in $O{P_n}$. In addition, the statistic has two degrees of freedom, which are $dof_1^{}$ and $dof_2^{}$.

\section{Decomposition Model Assisted DQN for Energy-Saving Decision Making}
In this section, we describe the realization of the decomposition model assisted energy-saving scheme. We introduce the proposed DQN-based method for choosing energy-saving operations and illustrate how the decomposition model assists the BSs agent in making decisions. 

\begin{algorithm}[t]
	\caption{Tukey's HSD} 
	\hspace*{0.02in} {\bf Input:} multiple energy-saving operations combinations \\ 
	\hspace*{0.02in} {\bf Output:} conflict or no conflict
	\begin{algorithmic}[1]
		\While {each episode} 
		
		\State{Select two combinations of energy-saving operations }
		\State{Calculate the difference in means ${\rm{m_{diff}}}$}
		\State{Calculate the standard errors $e_s$}
		\State{Calculate the Student's range statistic $q$}
		\State{Calculate the critical value $q_{th}^{} = q \times e_s$}
		\State{Identify conflicts
			$\left\{ {\begin{array}{*{20}{c}}
					{true,}&{{\rm{if\;\;m_{diff} > {q_{th}}}}},\\
					{false,}&{{\rm{otherwise, }}}
			\end{array}} \right.$}
		\EndWhile
	\end{algorithmic}
\end{algorithm}

\subsection{MDP-Guided DQN}\label{A} 
Our goal is to solve the established optimization problem and find the optimal combination of energy-saving operations. Markov decision processes (MDP) provide a theoretical framework for decision-making problems. Guided by this framework, the DQN approximates the Q-function using two neural networks to find the best decision in a complex high-dimensional state space and action space [26]. 

\subsubsection{Decision Time}\label{1)} $t \in \left[ {1,2, \cdots ,T} \right]$, at certain intervals, the system evaluates the current state, selects the optimal combination of  energy-saving operations, and executes them.

\subsubsection{State Space}\label{2)} the state space $S$ refers to three network objectives of the energy consumption, the throughput, and the first packet delay that we are considered, which is

\vspace{-0.2cm}
\begin{equation}
	\label{deqn_ex1}
	\setlength\abovedisplayskip{6pt}
	\setlength\belowdisplayskip{6pt}
	{S_{}} = \{ {s^1},{s^2},{s^3}\},
\end{equation}	
\vspace{-0.4cm}

\noindent
where ${s^1}$, ${s^2}$ and ${s^3}$ are the average of the energy consumption, the throughput, and the first packet delay, respectively, at the decision time $t$.

\subsubsection{Action Space}\label{3)} the action space $A$ refers to three types of energy-saving operations that the BSs agent can make, which is

\vspace{-0.4cm}
\begin{equation}
	\label{deqn_ex1}
	\setlength\abovedisplayskip{6pt}
	\setlength\belowdisplayskip{6pt}
	{A_{}} = \{ a_{}^1,a_{}^2,a_{}^3\},
\end{equation}	
\vspace{-0.4cm}

\noindent
where $a_{}^1$ is the adjustment of the transmit power $P_{Tx}^{}$, the possible values of $P_{Tx}^{}$ are $\{ 53,52,51,50,0\}$; $a_{}^2$ is the adjustment of the down-tilt angle ${\theta _{}}$, the possible values of ${\theta _{}}$ are $\{ 5_{}^ \circ ,15_{}^ \circ ,20_{}^ \circ ,25_{}^ \circ ,0_{}^ \circ \}$; $a_{}^3 = 1$ represents that the BSs are sleep, that is  $P_{Tx}^{}=0$, ${\theta _{}}=0_{}^ \circ$, and $a_{}^3 = 0$ represents that the BSs are active. The dimension of the action space is determined by the number of BSs.

\subsubsection{Reward Function}\label{4)} 
rewards are numerical values given to the action of the BSs agent to evaluate the accuracy of chosen actions. The goal of the BSs agent is to maximize long-term cumulative rewards. When the BSs agent takes an action that results in lower energy consumption, higher throughput, and lower first packet delay, the BSs agent will be rewarded positively. Conversely, it will be rewarded negatively. Therefore, we categorize network objectives into positive and negative objectives. For positive objectives, we want them to be as big as possible, e.g., the throughput, while for negative objectives, we want them to be as small as possible, e.g., the energy consumption and the first packet delay.

Therefore, we limit the contribution of each network objective to overall reward within the range $[0,1]$, and the reward function can be achieved by
\vspace{0.1cm}
\begin{equation}
	\label{deqn_ex1}
	\begin{array}{l}
		r = {w_1}\frac{{\rm{{\cal{\bar U}}}} - min (\rm{\cal{U}}^{THP})}{{max (\rm{\cal{U}}^{THP})} - {min (\rm{\cal{U}}^{THP})}}\\[10pt]
		\;\;\;- {w_2}\frac{{\rm{{\cal{\bar Z}}}} - min (\rm{\cal{Z}}^{EC})}{{max (\rm{\cal{Z}}^{EC})} - {min (\rm{\cal{Z}}^{EC})}} - {w_3}\frac{{\cal{\bar V}} - min (\rm{\cal{V}}^T)}{{max (\rm{\cal{V}}^T)} - {min (\rm{\cal{V}}^T)}},
	\end{array}
\end{equation}	
\vspace{0.1cm}

\noindent
where a set of the first packet delay data as $\rm{\cal{V}}_{}^{T} = [v_1^{},v_2^{}, \cdots ,v_L^{}]$ represents the average of this set of the first packet delay data. $max (\rm{\cal{U}}^{THP})$, $max (\rm{\cal{V}}^T)$ represent the maximum value of the throughput and the and the first packet delay, respectively. $min (\rm{\cal{U}}^{THP})$, $min (\rm{\cal{V}}^T)$ represent the minimum value of the throughput and the first packet delay, respectively. ${w_1}$, ${w_2}$, ${w_3}$ are the weights, which are adjusted to realize the trade-off among the energy consumption, the throughput, and the first packet delay.

\subsection{Algorithm Description}\label{B} 

As shown in Algorithm 2, we present the concrete implementation steps of the proposed decomposition model assisted DQN method. The decomposition model outputs combinations of energy-saving operations stored as JSON format, and we set up this JSON file as the action space of the DQN. The BSs agent utilize experience playback and phase update to explore effective energy-saving operations under this action space.

\begin{algorithm}[t]
	\caption{Decomposition Model Assisted DQN} 
	\hspace*{0.02in} {\bf Input:}\\ 
	\hspace*{0.22in}JSON file of energy-saving operations combinations\\
	\hspace*{0.22in}Initialize: simulation time, step size, action space\\
	\hspace*{0.22in}Initialize: $\forall s \in S,\forall a \in A, Q(s,a), D$, $\gamma\in (0,1]$\\
	\hspace*{0.02in} {\bf Output:}\\
	\hspace*{0.22in}The values of the energy consumption, the throughput, 	\hspace*{0.22in}and the first packet delay and the combinations of energy-\hspace*{0.22in}saving operations.\\
	\hspace*{0.02in} {\bf Function1: DQN make decisions}
	\begin{algorithmic}[1]
		\For {each episode} 
		\For{$t = 1,T$}  
		\State{Select ${a_t} = \arg {max _a}Q({s_t},a;\omega )$}, execute ${a_t}$
		\State{Set ${s_{t + 1}} = {s_t}$, store $({s_t},{a_t},{s_{t + 1}},{r_t})$ in $D$}
		\State{Extract minibatch $({s_j},{a_j},{s_{j + 1}},{r_j})$ from $D$}
		\State{Set ${y_j} = \left\{ {\begin{array}{*{20}{c}}
					{{r_j},\;{\rm{for\;terminal}}\;j + 1}\\
					{{r_j} + \gamma {{max }_{a'}}\widetilde Q({s_{j + 1}},a';{\omega ^ - }),{\rm{otherwise}}}
			\end{array}} \right.$}
		\State{Perform gradient descent ${({y_j} - Q({s_j},{a_j};\omega ))^2}$}
		\State{Update network parameters $\omega $, reset $\tilde Q = Q$}
		\EndFor
		\EndFor
	\end{algorithmic}
	\hspace*{0.02in} {\bf Function2: Decomposition model update}
	\begin{algorithmic}[1]
		\For {each episode}
		\State $LSGs \to SG$ weights update in (8)
		\State $OPs \to LSGs$ weights update in (9)
		\State The score of SG is calculated in [9]
		\If{$S{G_{score}} \le S{G_{th}}$}
		\State Identifying conflicts between network objectives
		\State Identifying conflicts between operations
		\State New energy-saving operations combinations
		\Else
		\State Pass
		\EndIf
		\EndFor
	\end{algorithmic}
\end{algorithm}

Specifically, at each time step $t$, the BSs agent initializes the simulation environment and returns the current state ${s_t}$, ${s_t} \in S$. The BSs agent then selects the optimal combination of energy saving options based on ${a_t} = \arg {max _a}Q({s_t},a;\omega )$, ${a_t} \in A$, and executes them. Subsequently, the state ${s_t}$ transitions to the state ${s_{t + 1}}$, meanwhile the BSs agent receives reward $r_t^{}$, and stores the $({s_t},{a_t},{s_{t + 1}},{r_t})$ in replay memory $D$. The BSs agent maintains two Q-networks, mainly includes target network and evaluate network. The target network is used to update the parameter $\omega$ of the evaluate network, and the evaluate network is used to calculate the expected reward for each possible action. The BSs agent uses bellman equation to evaluate the weights, expressed by

\vspace{-0.1cm}
\begin{equation}
	\label{deqn_ex1}
	{y_j} = {r_j} + \gamma {max _{a'}}\widetilde Q({s_{j + 1}},a';{\omega ^ - }),
	\vspace{-0.2cm}
\end{equation}	
\vspace{-0.1cm}

\noindent
where ${y_j}$ is the Q value at step $j$,  $\gamma$ is the discount factor, $\gamma\in (0,1]$, and $\widetilde Q({s_{j + 1}},a';{\omega ^ - })$ is the Q value of the target network.

The method of gradient descent is employed to update the value of $\omega$, which is calculated by 

\vspace{0.0cm}
\begin{equation}
	\label{deqn_ex1}
	Loss\left( \omega  \right) = E\left[ {{{\left( {{y_j} - Q({s_j},{a_j};\omega )} \right)}^2}} \right],
\end{equation}	
\vspace{-0.2cm}

\noindent
where $Q({s_j},{a_j};\omega )$ is the Q value of the evaluate network. 

We pass the output data to the decomposition model at the end of one training session of the DQN. Then, we update the weights, calculate the score of SG and denote it by $S{G_{score}}$, which reflects the fulfillment of the energy-saving intent. If $S{G_{score}}$ is lower than the threshold $S{G_{th}}$, conflict identifications performed and finally output new energy-saving operations combinations.

\section{Simulation Results}

In this section, we provide simulation results to evaluate the effectiveness of our proposed decomposition model assisted energy-saving scheme. We consider a wireless communication scenario with $N=4$ BSs and $M=32$ UEs in a cluster. We use python programming language to build simulation platform, and the simulation parameters are set based on the urban Macro BSs model of 3GPP TR38.912 [27]. The detailed simulation parameters are shown in Table I. The simulation platform mainly contains three python classes, ``User class'', ``BS class'', ``Channel class'', and each class contains some attributes and function capabilities. For example, the relevant attributes in the ``User class'' include UE's location, UE's service arrival rate and UE's index, etc. And the relevant attributes in the ``BS class'' include BS's transmit power, BS's antenna gain and BS's index, etc. Functions can perform actions in the wireless communication simulation platform. For example, functions in the ``User class'' include the generation of UE's communication service, the change of UE's location and the selection of UE's cell, etc. Functions in the ``BS class'' the schedule of includes BS's resource, the allocation of power, the determination of MCS and the update of BS's buffer. As an intermediate channel connecting UEs and BSs, the ``Channel class'' mainly carries out some measurements and calculations of channel conditions. Functions in the ``Channel class'' include the calculation of reference signal receiving power and the determination of the information transmission results, etc. In addition, we use simpy to realize discrete event simulation and the queuing behavior of transmission service. The training process is 1000 steps per episode, and each training step represents 100 TTIs. The overall process mainly consists of two parts:

\begin{table}
	\renewcommand\arraystretch{1.3}
	\begin{center}
		\caption{Simulation Parameters.}
		\label{tab1}
		\begin{tabular}{|l|l|}
			\hline
			Parameter & Value\\
			\hline
			\hline
			Number of Sectors ${N_s}$ & 3\\
			\hline
			Altitude $H$ & 25m\\ 
			\hline
			System bandwidth $B$ & 10/20/40/100MHZ\\
			\hline 
			Number of RBs $r_n^{\max }$ & 52/106/216/273\\
			\hline
			System frequency $f_c^{}$ & 3.5GHz\\
			\hline
			RB Bandwidth ${B_{prb}}$ & 180KHz\\
			\hline
			Arrival rate $\lambda $ & 1/2/4/8\\
			\hline
			Transmit Power $P_{Tx}^n$ & 53/52/51/50dBm\\
			\hline
			constant ${\xi _i}$ & 21.45\\
			\hline
			constant ${\psi _i}$ & 354.44\\
			\hline
			Downtilt Angle ${\theta _{}}$ & ${0^ \circ }/ {5^ \circ }/ {15^ \circ }/ {20^ \circ }$\\
			\hline
			Noise power spectral density $\sigma _{}^2$ & -174dBm/Hz\\
			\hline
			SINR Threshold ${p_0}$ & -5.1dBm\\
			\hline
		\end{tabular}
	\end{center}
	\vspace{-0.4cm}
\end{table}

\begin{itemize}
	\item[$\bullet$]Scheduling Process: at each TTI, computing the reference signal receiving power of each UE, and determining the MCS to schedule the RBs. Then the service packets are transmitted, and the buffer of each BS also is updated, recording the data of three network objectives.
	
	\item[$\bullet$]Decision Making: at each training step, the average of the energy consumption $\bar P$, the average of the throughput $\bar C$, the average of the first packet delay $\bar T$, and the reward $r$ are calculated. The BSs agent adjusts energy-saving operations according to the reward $r$, and fills replay memory $D$. Meanwhile, the UEs reselect the BSs for connection. Particularly, the BSs agent is trained, and network parameters are updated when time step is greater than 200 steps.
\end{itemize}

\subsection{Decomposition Model Assisted Scheme Performance}\label{A}

We test the adaptive adjustment performance of the decomposition model assisted scheme when the number of network elements changes. We set $N=10$ BSs and $M=80$ UEs. During the run time, the trained DQN is loaded to select a combination of energy-saving operations for each BSs. And then, we pass the set of data output from the run time to the design time and test the time overhead of the whole design time. As shown in Fig. 4, we test the execution time several times, the result shows that the average of the execution time is $1.03 s$, which means that the adaptive adjustment time is less than $2 s$. Therefore, the decomposition model assisted scheme has a good adjustment performance when the number of BSs and UEs changes.

\begin{figure}[!t]
	\centering
	\includegraphics[width=3.6in]{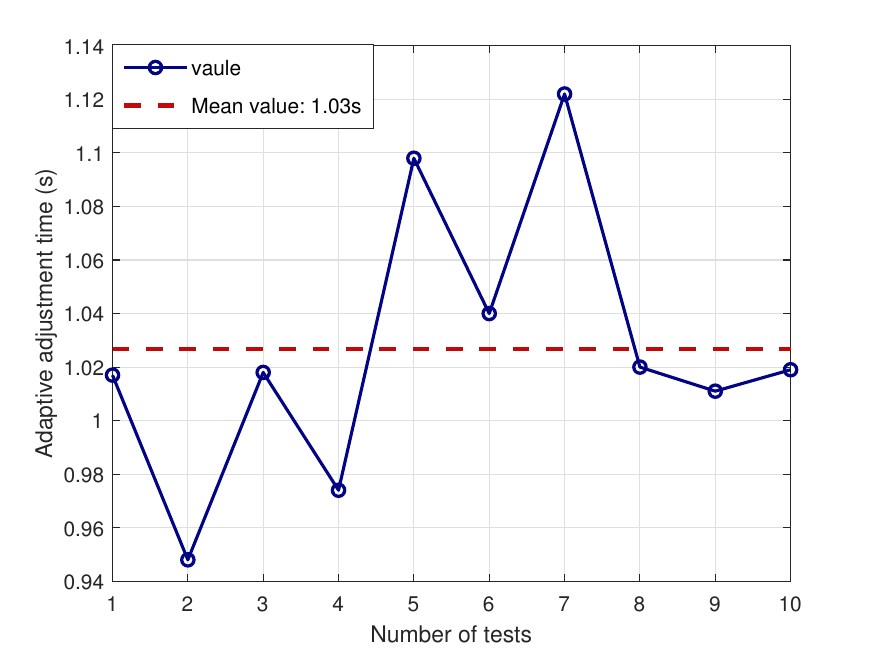}
	\caption{The adaptive adjustment time testing of the decomposition model assisted scheme.}
	\label{Fig4}
	\vspace{-0.45cm}
\end{figure}

In Fig. 5, we present the trend of the loss function for the DQN. Fig. 5(a) applies our proposed decomposition model assisted scheme, Fig. 5(b) does not apply our proposed decomposition model assisted scheme. The result shows that when the decomposition mode of the design time reduces the action space dimension of the DQN, the learning process is more stable and the model shows better convergence. This is due to the fact that a smaller action space dimension reduces the complexity of the search space, making the impact of each action more significant and mitigating the unstable learning dynamics due to function approximation using neural networks. Furthermore, it is obvious that our proposed scheme makes the value of loss function smaller. This is due to the fact that the size of the action space directly affects the distribution of Q-values. In a larger action space, there may be more actions that are close to the optimum, which may lead to finer differences in Q-values. Conversely, in a smaller action space, the difference in Q-values for each action may be more significant, which leads to a larger absolute value of the loss function. 

Then, we compare the change in the cumulative reward of the DQN, as illustrated in Fig. 6. The trend of the cumulative reward shows an initial decrease, reflecting the exploration process, followed by a subsequent increase in logarithmic form, indicating that the algorithm is gradually finding a more effective strategy. Fig. 6(a) applies our proposed decomposition model assisted scheme, the cumulative reward starts to enter the increase phase at $step=260$. Fig. 6(b) does not apply our proposed decomposition model assisted scheme, the cumulative reward starts to enter the increase phase at $step=160$. This suggests that the size of the action space affects the speed of learning and the number of explorations required. When the decomposition model of the design time reduces the action space, the learning process can be accelerated so that the cumulative reward enters the increase phase more quickly.

\begin{figure}
	\centering
	\subfigure[Proposed decomposition model assisted scheme.]{
		\begin{minipage}[b]{0.382\textwidth}
			\includegraphics[width=1\textwidth]{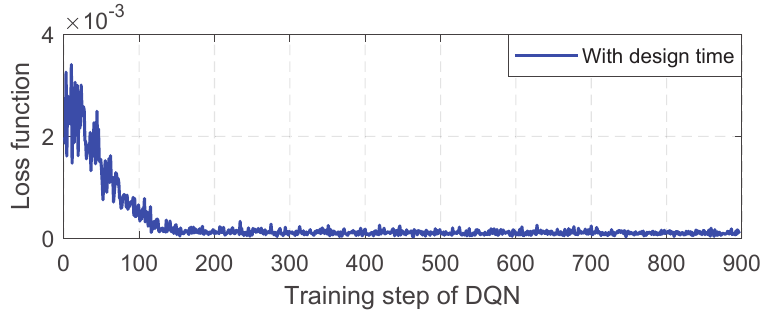}
		\end{minipage}
		\label{Simulation_Results_4_a}
	}
	\subfigure[Without decomposition model assisted scheme.]{
		\begin{minipage}[b]{0.382\textwidth}
			\includegraphics[width=1\textwidth]{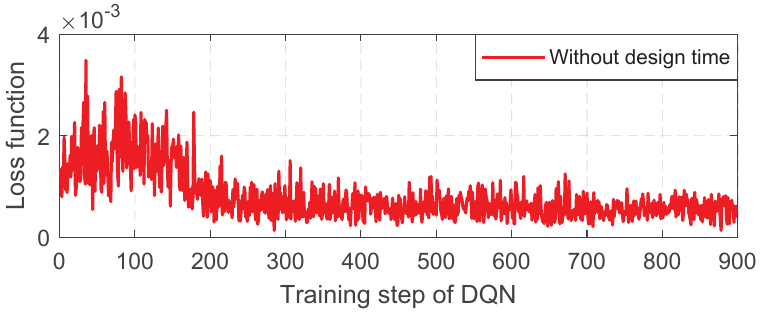}
		\end{minipage}
		\label{Simulation_Results_4_b}
	}
	\caption{The effect of the decomposition model on DQN loss function.}
	\label{Simulation_Results_4}
	\vspace{-0.3cm} 
\end{figure}

\begin{figure}
	\centering
	\subfigure[Proposed decomposition model assisted scheme.]{
		\begin{minipage}[b]{0.382\textwidth}
			\includegraphics[width=1\textwidth]{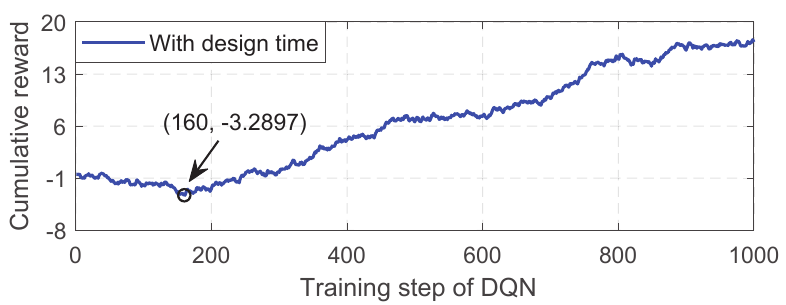}
		\end{minipage}
		\label{Simulation_Results_4_a}
	}
	\subfigure[Without decomposition model assisted scheme.]{
		\begin{minipage}[b]{0.382\textwidth}
			\includegraphics[width=1\textwidth]{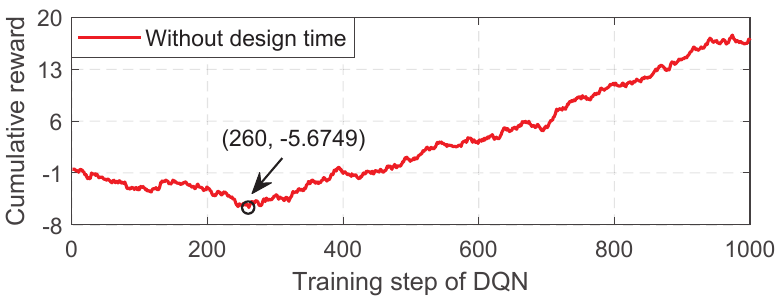}
		\end{minipage}
		\label{Simulation_Results_4_b}
	}
	\caption{The effect of the decomposition model on DQN cumulative reward function.}
	\label{Simulation_Results_4}
	\vspace{-0.3cm} 
\end{figure}

\subsection{Decomposition Model Assisted Scheme on Energy-Saving Performance}\label{B}

The variations in the action space of the DQN significantly affect its decision results. Therefore, in Fig. 7, we show the performance of three schemes on three network objectives. In Fig. 7(a), we illustrate the performance in the average energy consumption among our proposed decomposition model assisted scheme, the scheme with decomposition model assisted but without conflict identification, and the scheme without decomposition model assisted. In contrast, our proposed decomposition model assisted scheme has the lowest energy consumption. In Fig. 7(b), we show the impact of the three schemes on the average downlink throughput, it is obvious that our proposed decomposition model assisted scheme is able to achieve higher throughput. And in Fig. 7(c), we show the impact of the three schemes on the average first packet delay, it is obvious that our proposed decomposition model assisted scheme is able to achieve lower delay. Each scheme corresponds to a different dimension of the  DQN action space. Our proposed decomposition model assisted scheme is able to improve the throughput and reduce the first packet delay while reducing energy consumption compared to the scheme without decomposition model assisted. This finding confirms the effectiveness of decomposition model assisted DQN decisions making in balancing multiple network objectives, thereby improving the overall network performance. In addition, our decomposition model assisted scheme shows higher effectiveness in performing conflict identification compared to scheme that do not perform conflict identification.

\begin{figure*}[htb]
	\vspace{-0.3cm}
	\centering
	\subfigcapskip=-6.1pt
	\subfigure[]{
		\begin{centering}
			\includegraphics[scale=0.4]{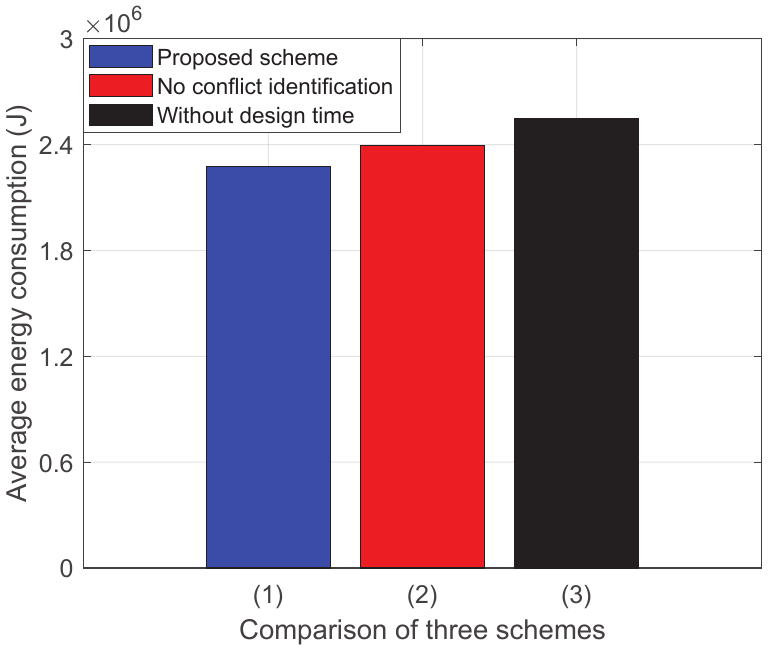}
		\end{centering}\hspace{-4mm}
		\label{Fig4a}}
	\subfigure[]{
		\begin{centering}
			\includegraphics[scale=0.4]{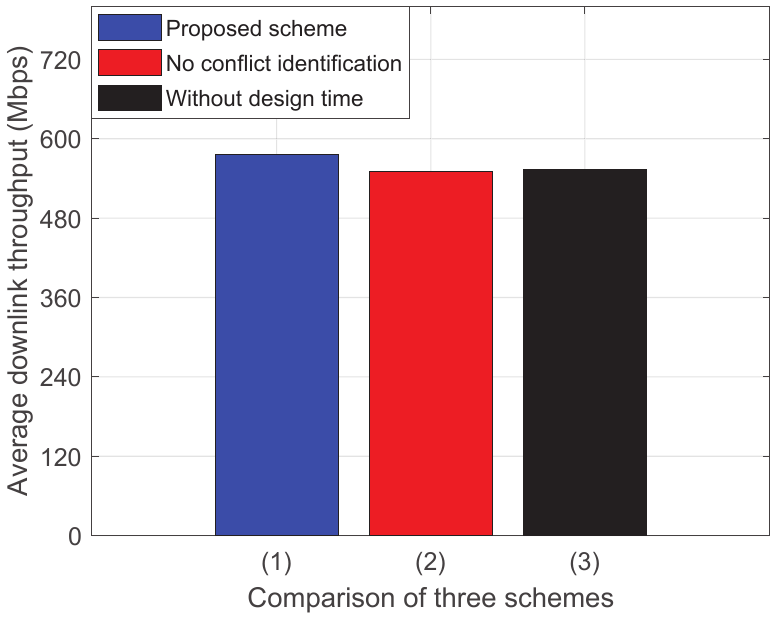}
		\end{centering}\hspace{-4mm}
		\label{Fig4b}}
	\subfigure[]{
		\begin{centering}
			\includegraphics[scale=0.4]{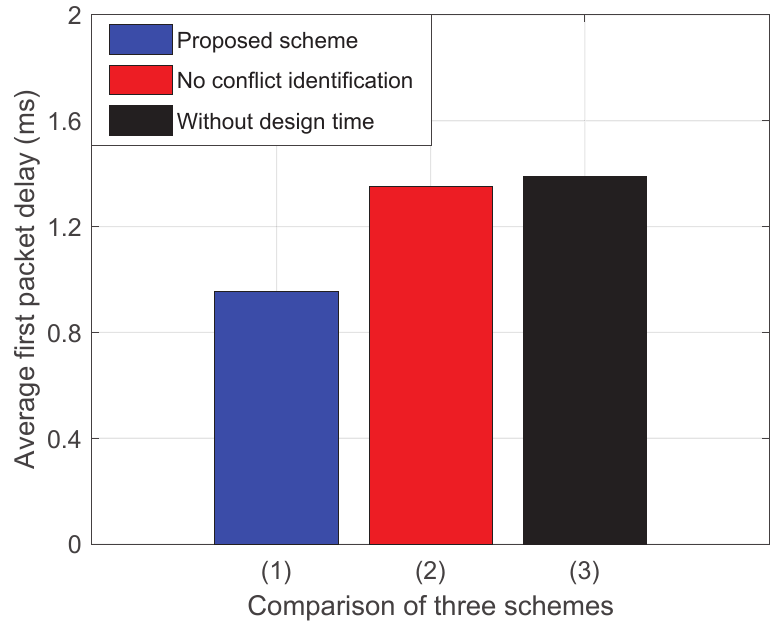}
		\end{centering}
		\label{Fig4c}}
	\vspace{-0.3cm}
	\caption{The performance of three schemes on the average energy consumption, the average downlink throughput, the average first packet delay.}
	\label{fig2}
	\vspace{-0.48cm}
\end{figure*}

\section{Conclusion}
6G networks will provide extensive network capacity and widespread fast access through ultra-dense deployment of BSs, while achieving significant improvements in higher bandwith and ultra-low delay. However, this also lead to an increase in energy consumption, thus research on reducing energy consumption has become crucial. In this work, we proposed a novel decomposition model assisted energy-saving design scheme that enabled assistance to the DQN in making energy-saving decisions and effectively traded-off among the energy consumption, the throughput, and the first packet delay. In addition, we utilized two methods for updating weights of the decomposition model to accurately measure the fulfillment of the energy-saving intent. We also utilized two methods for conflicts identification to influence the output of the decomposition model. Moreover, we evaluated the performance of the decomposition model assisted energy-saving scheme through simulation, and the results showed that the scheme is effective in energy-saving.


\begin{thebibliography}{100}
	\bibitem{b1} 
	T. Islam, D. Lee and S. S. Lim, ``Enabling network power savings in 5G-advanced and beyond,'' \emph{IEEE J. Sel. Areas Commun.}, vol. 41, no. 6, pp. 1888--1899, Jun. 2023.
	
	\bibitem{b2}
	T. Yu, S. Zhang, X. Chen and X. Wang, ``A novel energy efficiency metric for next-generation green wireless communication network design,'' \emph{IEEE Internet Things J.}, vol. 10, no. 2, pp. 1746--1760, Jan. 2023.
	
	\bibitem{b3}
	A. Havolli and M. Fetaji, ``Improving radio network planning and design in next-generation mobile networks using AI and ML algorithms,'' in \emph{Proc. 12th Mediterranean Conf. Embed. Comput. (MECO)}, Budva, Montenegro, Jun. 2023.
	
	\bibitem{b4}
	R. ElKazzi and A. Khalil, ``Energy-saving solution for future cellular systems,'' in \emph{Proc. 4th Int. Conf. on Renew. Energ. Dev. Ctries. (REDEC)}, Beirut, Lebanon, Nov. 2018.
	
	
	\bibitem{b5new1}
	K. C. Chang, K. C. Chu, H. C. Wang, Y. C. Lin and J. S. Pan, ``Energy saving technology of 5G base station based on internet of things collaborative control,'' \emph{IEEE Access}, vol. 8, pp. 32935-32946, Feb. 2020.
	
	\bibitem{b6}
	S. E. Nwachukwu, M. Chepkoech, A. A. Lysko, K. Awodele, J. Mwangama and C. R. Burger, ``Integration of massive MIMO and machine learning in the present and future of power consumption in wireless networks: A review,'' in \emph{Proc. IEEE 7th Forum Research Technol. Soc. Ind. Innov. (RTSI)}, Paris, France, Aug. 2022, pp. 154--160.
	
	\bibitem{b7}
	M. E. Morocho-Cayamcela, H. Lee and W. Lim, ``Machine learning for 5G/B5G mobile and wireless communications: Potential, limitations, and future directions,'' \emph{IEEE Access}, vol. 7, pp. 137184--137206, Sep. 2019
	
	\bibitem{b8new3}
	T. Rumeng, W. Tong, S. Ying and H. Yanpu, ``Intelligent energy saving solution of 5G base station based on artificial intelligence technologies,'' in \emph{Proc. IEEE International Joint EMC/SI/PI and EMC Europe Symposium}, Raleigh, NC, USA, Oct. 2021, pp. 739-742.
	
	
	\bibitem{b9new4}
	J. Yuan, L. Zhang, T. Lv and Y. Wu, ``A method for improving the effect of base station energy saving with AI,'' in \emph{Proc. IEEE 23rd Int Conf on High Performance Comput.  Commun. (HPCC)},  Haikou, Hainan, China, May. 2021, pp. 2060-2065.
	
	
	\bibitem{b10}
	Y. Azimi, S. Yousefi, H. Kalbkhani and T. Kunz, ``Applications of machine learning in resource management for RAN-slicing in 5G and beyond networks: A survey,'' \emph{IEEE Access}, vol. 10, pp. 106581--106612, Sep. 2022.
	
	
	\bibitem{b11}
	F. E. Salem, Z. Altman, A. Gati, T. Chahed and E. Altman, ``Reinforcement learning approach for advanced sleep modes management in 5G networks,'' in \emph{Proc. IEEE 88th Veh. Technol. Conf. (VTC-Fall)}, Chicago, IL, Aug. 2018.
	
	
	\bibitem{b12}
	M. Javad-Kalbasi, Z. Naghsh, M. Mehrjoo and S. Valaee, ``A new heuristic algorithm for energy and spectrum efficient user association in 5G heterogeneous networks,'' in \emph{Proc. IEEE 31st Ann. Int. Symp. Person. Indoor Mob. Radio Commun.}, London, UK, Aug. 2020.
	
	
	\bibitem{b13}
	C. Luo and J. Liu, ``Load based dynamic small cell on/off strategy in ultra-dense networks,'' in \emph{Proc. 10th Int. Conf. Wirel. Commun. Signal Process. (WCSP)}, Hangzhou, China, Oct. 2018.
	
	
	\bibitem{b14}
	M. Masoudi, M. G. Khafagy, E. Soroush, D. Giacomelli, S. Morosi and C. Cavdar, ``Reinforcement learning for traffic-adaptive sleep mode management in 5G networks,'' in \emph{Proc. IEEE 31st Ann. Int. Symp. Person. Indoor Mob. Radio Commun.}, London, UK, Sep. 2020.
	
	
	\bibitem{b15}
	A. E. Amine, J. -P. Chaiban, H. A. H. Hassan, P. Dini, L. Nuaymi and R. Achkar, ``Energy optimization with multi-sleeping control in 5G heterogeneous networks using reinforcement learning,'' \emph{IEEE Trans. Netw. Serv. Manag.}, vol. 19, no. 4, pp. 4310--4322, Dec. 2022.
	
	
	\bibitem{b16}
	Q. Zhao, S. Paris, T. Veijalainen and S. Ali, ``Hierarchical multi-objective deep reinforcement learning for packet duplication in multi-connectivity for URLLC,'' in \emph{Proc. Jt. Eur. Conf. Netw. Commun. 6G Summit (EuCNC/6G Summit)}, Porto, Portugal, Jun. 2021, pp. 142--147.
	
	
	\bibitem{b17}
	F. E. Salem, Z. Altman, A. Gati, T. Chahed and E. Altman, ``Reinforcement learning approach for advanced sleep modes management in 5G networks,''in \emph{Proc. IEEE 88th Veh. Technol. Conf. (VTC-Fall)}, Chicago, IL, Aug. 2018.
	
	\bibitem{b18}
	S. Malta, P. Pinto and M. Fernández-Veiga, ``Using reinforcement learning to reduce energy consumption of ultra-dense networks with 5G use cases requirements,'' \emph{IEEE Access}, vol. 11, pp. 5417--5428, Jan. 2023.
	
	
	\bibitem{b19}
	S. K. Gowtam Peesapati, M. Olsson and S. Andersson, ``A multi-strategy multi-objective hierarchical approach for energy management in 5G networks,'' in \emph{Proc. IEEE Global Commun. Conf. (GLOBECOM)}, Rio de Janeiro, Brazil, Dec. 2022, pp. 4842--4847.
	
	
	\bibitem{b20}
	M. F. Hossain, K. S. Munasinghe and A. Jamalipour, ``Toward self-organizing sectorization of LTE eNBs for energy efficient network operation under QoS constraints,'' in \emph{Proc. IEEE Wirel. Commun. Netw. Conf. (WCNC)}, Shanghai, China, Apr. 2013, pp. 1279--1284.
	
	\bibitem{b21}
	Y. Liao, Z. Yang, Z. Yin and X. Shen, ``DQN-based adaptive MCS and SDM for 5G massive MIMO-OFDM downlink,'' \emph{IEEE Commun. Lett.}, vol. 27, no. 1, pp. 185--189, Jan. 2023.
	
	
	\bibitem{b22}
	K. Zhang, R. Zhang, J. Wu, Y. Jiang, and X. Tang, ``Measurement and modeling of path loss and channel capacity analysis for 5G UMa scenario,'' in \emph{Proc. 11th Int. Conf. Wirel. Commun. Signal Process. (WCSP)}, Xi'an, China, Oct. 2019.
	
	
	\bibitem{b23}
	E. J. Scheid, C. C. Machado and M. F. Franco, ``INSpIRE: integrated nfv-based intent refinement environment,'' in \emph{Proc. IFIP/IEEE Symposium on Integrated Netw. and Serv. Manag. (IM)}, Lisbon, Portugal, pp. 186-194, May. 2017.
	
	
	
	\bibitem{b23}
	S. Wang and L. Zhang,``A Supervised Correlation Coefficient Method: Detection of Different Correlation,'' in \emph{Proc. 12th Intl. Conf. on Advanced Comput. Intelligence (ICACI)}, Dali, China, Aug. 2020.
	
	
	
	\bibitem{b24}
	A. Nanda, A. P. K. Mahapatra and B. B. Mohapatra, ``Multiple comparison test by Tukey's honestly significant difference (HSD): Do the confident level control type I error,'' \emph{Int. J. Appl. Math. Stat.}, vol. 6, no. 1, pp. 59--65, Jun. 2021.
	
	
	\bibitem{b25}
	L. Liu, K. Xiong, Y. Lu, P. Fan and K. B. Letaief, ``Age-constrained energy minimization in UAV-assisted wireless powered sensor networks: A DQN-based approach,'' in \emph{Proc. IEEE Conf. Comput. Commun. Wkshp. (INFOCOM WKSHPS)}, Vancouver, BC, Canada, May 2021.
	
	\bibitem{b26}
	\textit{Study on enhanced intent driven management services for mobile networks.} Accessed: Jun. 2023. [Online]. Available: https://portal.3gpp.org/desktopmodules/Specifications/Specification\\Details.aspx?specificationId=3969
	
\end{thebibliography}
\end{document}